\documentclass[twocolumn]{aastex701}

\usepackage{xcolor}
\usepackage{gensymb}

\usepackage{amsmath}
\usepackage{bm}

\begin{document}

\title{Electromagnetic Precursors to Binary Neutron Star Mergers: Kinetic Simulations of Magnetospheric Flaring}

\author[0000-0002-6013-4655]{Jasmine Parsons}
\affiliation{Department of Astrophysical Sciences, Princeton University, 4 Ivy Lane, Princeton, NJ 08540}
\email{jasmine.parsons@princeton.edu}

\author[0000-0001-9179-9054]{Anatoly Spitkovsky}
\affiliation{Department of Astrophysical Sciences, Princeton University, 4 Ivy Lane, Princeton, NJ 08540}
\email{anatoly@princeton.edu}

\author[0000-0001-7801-0362]{Alexander Philippov}
\affiliation{Department of Physics, University of Maryland, 7901 Regents Drive, College Park, MD 20742, USA}
\email{sashaph@umd.edu}

\author[0000-0001-8939-6862]{Hayk Hakobyan}
\affiliation{Center for Computational Astrophysics, Flatiron Institute, 162 Fifth Ave., New York, NY 10010, USA}
\email{haykh.astro@gmail.com}

\submitjournal{ApJL}

\begin{abstract}

We present the first 3D global kinetic simulations of the interacting magnetospheres of pre-merger binary neutron stars. The stars, whose magnetic moments are anti-aligned, twist the field lines connecting them, leading to periodic eruptions. Each eruption consists of an expanding magnetic flux tube with a reconnecting current sheet trailing behind it, topologically analogous to coronal mass ejections. We predict two novel classes of electromagnetic precursor signals powered by the efficient dissipation of magnetic energy in these periodically forming trailing current sheets. First, particles accelerated in the sheets produce nonthermal gamma-ray signals peaking at $\sim16\,\mathrm{MeV}$, which escape minutes to seconds before merger while the sheets are still optically thin to pair production, with modest characteristic luminosities of $L_\mathrm{obs}\gtrsim10^{42}\,\mathrm{erg/s}$, detectable only for nearby mergers. Second, merging plasmoids in the sheets could produce fast radio burst-like transients in the final seconds before merger, with characteristic luminosities $L_\mathrm{radio}\sim10^{38-40}\,\mathrm{erg/s}$. These coherent radio precursors would be detectable by upcoming instruments, either in untargeted surveys by wide-field instruments such as CHORD, or through targeted follow-up of gravitational-wave early-warning alerts with instruments such as DSA or SKA-mid.

\end{abstract}


\section{Introduction} \label{sec:intro}

The landmark gravitational wave detection of the binary neutron star (BNS) merger GW170817 \citep{gw170817} and its subsequent multi-wavelength counterparts \citep[e.g.,][]{gw170817gammaray, coulter2017, pian2017} established a new paradigm for multi-messenger astrophysics. While postmerger signatures such as short gamma-ray bursts and kilonovae have been extensively modeled, there has been increasing attention in recent years to the possibility of electromagnetic emission preceding the merger itself. Such a precursor signal, generated by crustal or magnetospheric processes, would offer a direct probe of the physical conditions in these binary systems, while also potentially enabling early localization of impending post-merger electromagnetic signals.

Magnetospheric emission can arise because, as the stars inspiral, their relative orbital motion induces electric fields strong enough to reignite pair production, filling the magnetosphere with pair plasma capable of powering electromagnetic radiation \citep{lyutikov2019}, analogous to emission from pulsars and magnetars \citep{philippov2022review, magnetarreview}. Both analytic \citep{hansen2001, lai2012, lyutikov2019, zhangbing2020, cherkis2021, lyutikov2023} and numerical studies, using MHD \citep{paluenza2013, paluenza2013b, sharma2026}, force-free \citep{most2020apjl,most2022mnras, mostPRL, mahlmann2025, skiathas2025}, and 2.5D axisymmetric kinetic \citep{crinquand2019} simulations have demonstrated that pre-merger magnetospheres can efficiently dissipate energy in the final milliseconds before coalescence, for a variety of binary configurations.

In particular, the force-free simulations of \cite{most2020apjl} showed that, if the stars have anti-aligned dipoles, twist builds up on the field lines connecting the stars before being explosively released in periodic eruptions topologically analogous to solar coronal mass ejections (CMEs) \citep{most2020apjl, most2022mnras, mostPRL}. However, while force-free models can capture large-scale energy release, they are fundamentally limited by their inability to resolve the kinetic microphysics of magnetic reconnection and the resulting nonthermal particle acceleration. Kinetic simulations are thus required to bridge the gap between global magnetospheric dynamics and observable emission.

In this Letter, we present the first global 3D particle-in-cell (PIC) simulations of pre-merger BNS magnetospheres. We find that the CME-like eruptions identified in previous force-free work persist in the kinetic regime, and that the trailing current sheet behind each eruption is a site of efficient nonthermal particle acceleration. We propose that this trailing current sheet could power two novel electromagnetic precursor signatures which would occur minutes to seconds before merger. The first is a repeating nonthermal gamma-ray signal peaking at $\sim$16 MeV, and the second is a series of FRB-like coherent radio transients from merging plasmoids in the sheet. These repeating coherent radio precursors, in particular, could be detectable by upcoming instruments, either in untargeted wide-field surveys or through targeted follow-up of gravitational-wave early-warning alerts.

\section{Simulation Setup} \label{sec:setup}

We use the particle-in-cell code \texttt{TRISTAN-v2} \citep{hakobyan2024tristanv2}. Our domain is a uniform Cartesian grid of size $(L_x, L_y, L_z) = (560, 560, 880)\Delta x$, where $\Delta x$ is the cell size. Two stars of radius $R=20\Delta x$ are placed at a separation of $d = 70\Delta x$ along the x-axis, centered in $y$ and $z$. Each star is modeled as a perfectly conducting rotating sphere with a dipolar magnetic field imposed as a boundary condition, following previous global kinetic simulations of isolated pulsars \citep[see, e.g.,][]{philippov-oblique-pulsar, hakobyan2023pulsar}. The axis of rotation for both stars is along $z$, with one star having an inclination of $120^\circ$ and the other an inclination of $60^\circ$, so that their magnetic moments are anti-aligned. At the outer boundaries, electromagnetic fields are damped to zero, and particles are allowed to leave the simulation box.

The stars are fixed on the grid and rotate with spin period $P_* = 5000\Delta t$, where $\Delta t$ is the simulation timestep. In an inspiraling binary, the twist in the connecting field lines can be provided either by the orbital motion or by the stellar spins; in our simulation, we fix the stars on the grid for computational simplicity, and twist is instead supplied solely by the stellar rotation, which produces the same buildup of twist between the stars. Note that an isolated pulsar with this spin period would have a light cylinder radius $R_{LC} = c/\Omega_* \sim 358\,\Delta x$, outside of our simulation domain. However, the presence of the companion star compresses each magnetosphere (in a manner similar to \cite{zhong2024}), and so the Y-points of the individual stars always lie within the simulation box.

We fill the joint magnetosphere with pair plasma using an injection scheme designed to approach a nearly force-free solution (i.e., where $\bm{E}\cdot\bm{B} = 0$) \citep{goldreich1969}. In every cell where the normalized unscreened parallel electric field exceeds \mbox{$\bm{E}\cdot\bm{B}/B^2 > 0.015$}, we inject pairs with weight proportional to the unscreened field. We do not inject if the magnetization $\sigma = B^2/4\pi\rho c^2$ (where $B$ is the magnetic field strength and $\rho$ is the density) in the cell is below a threshold value of $\sigma_\mathrm{thresh} = 10$. Our results are insensitive to the exact values of these thresholds, as long as sufficient plasma is supplied such that the pair multiplicity $\kappa\equiv n_{e^{\pm}}/n_{\mathrm{GJ}}$ is maintained well above unity throughout the magnetosphere, where $n_\mathrm{GJ} = |\bm{\Omega}\cdot\bm{B}|/(2\pi ec)$ is the Goldreich-Julian number density \citep{goldreich1969}. We do not include radiative cooling. In addition, following previous global PIC work \citep[e.g.,][]{philippov-oblique-pulsar, hakobyan2023pulsar}, we ensure that the plasma skin depth is resolved ($d_e\gtrsim1$) throughout the domain, including near the equatorial and trailing current sheets, in order to properly capture reconnection physics.

\begin{figure*}[t] % or [!t] or [h] depending on where you want it
    \centering
    % 0.8 * page width in a two-column document
    \includegraphics[width=0.99\linewidth]{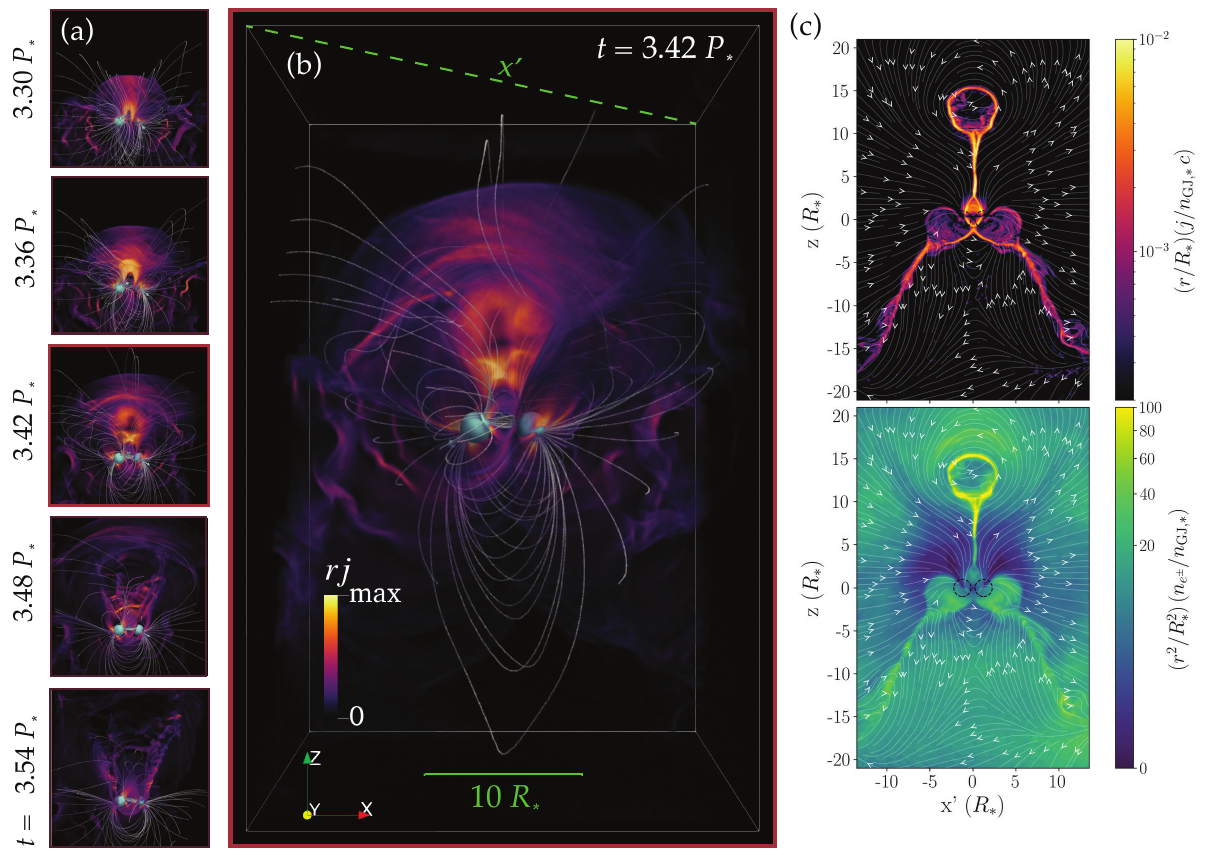}
    \caption{Three-dimensional rendering of our simulation of a pre-merger binary neutron star system. The main panel (b) shows a three-dimensional rendering of the current density $j$ (compensated by $r$) at $t = 3.42\,P_*$. The stars are rendered as light blue spheres, with arrows denoting their magnetic moments, and a sample of white field lines traced from their surfaces. Panel (a) shows five snapshots of burst evolution, from formation (top) to ejection (bottom). Panel (c) shows 2D slices of current density (top) and plasma density (bottom), on a plane perpendicular to the bubble and current sheet, passing midway between the two stars (indicated by the dashed green line in panel (b)). Dashed black lines in panel (c) denote the projection of the stars onto this plane. White field lines also show the magnetic field reversal across the current sheet. An animation showing the full simulation for multiple $P_*$ is available at \url{https://youtu.be/c8qWrmMkohQ}.}
    \label{fig:slice2d}
\end{figure*}

\begin{figure*}[t] % or [!t] or [h] depending on where you want it
    \centering
    % 0.8 * page width in a two-column document
    \includegraphics[width=0.7\paperwidth]{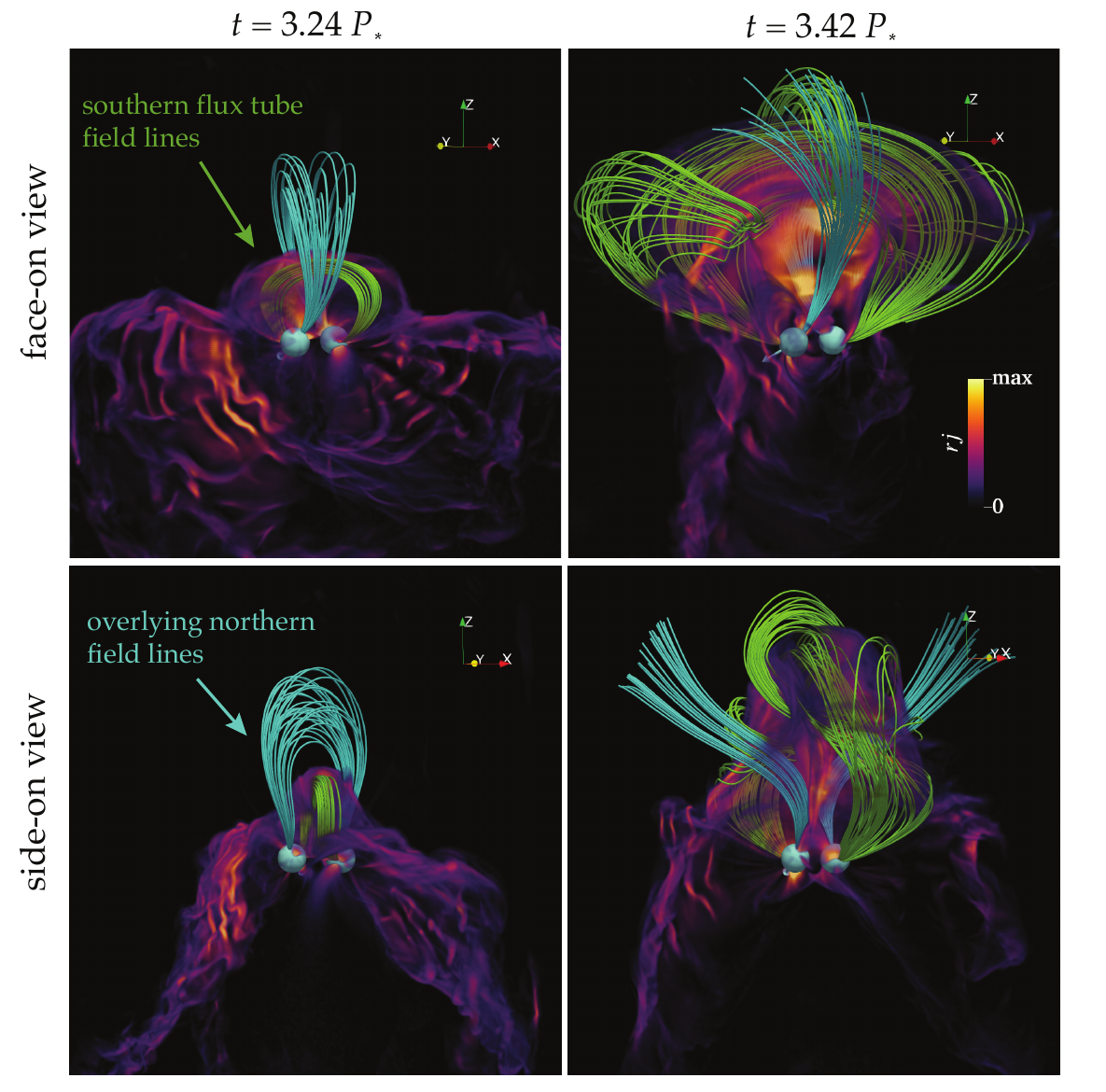}
    \caption{Volume rendering of the current density $j$ (compensated by $r$) at $t = 3.24\,P_*$ (left) and $t = 3.42\,P_*$ (right), shown face-on (top) and side-on (bottom) to the forming current sheet. Field lines of the rising southern flux tube are shown in green, and the overlying northern arcade in blue. At the earlier time, the flux tube has become perpendicular to the overlying arcade, analogous to pre-eruption conditions in solar CMEs. By the later time, the bubble has risen and the blue field lines are reconnecting behind it, forming the trailing vertical current sheet. Helical field lines in green can be seen in the top right panel, wrapping around the flux tube. An animation showing the full evolution of the burst is available at \url{https://youtu.be/E7Ch6GtBSXI}.
}
    \label{fig:fieldlines}
\end{figure*}

\section{Simulation Results} \label{sec:simresults}

\subsection{Global Magnetospheric Structure} \label{subsec:global}

Since the magnetic moments of the stars are anti-aligned, there are two sets of field lines that connect them: ones that connect the northern poles (`northern field lines') and ones that connect the southern poles (`southern field lines').\footnote{Note that when we refer to the `north' and `south' poles of the stars, we are referring to the top and bottom hemispheres of the stars relative to the simulation box, regardless of the direction of their magnetic moment.} As the stars rotate, twist builds up in these connecting field lines and is periodically released  in solar flare-like bursts. Each burst is composed of a rising magnetic flux rope, or `bubble', with a reconnecting current sheet trailing behind it, topologically analogous to the standard picture of CMEs. We call this trailing sheet the `vertical current sheet,' to distinguish it from the equatorial current sheets of the individual stars, seen in isolated pulsars \citep{spitkovsky2006, chen_2014, philippov-oblique-pulsar, hakobyan2023pulsar}. Since the stars are inclined, the magnetic poles of the stars are periodically brought closer together and farther apart. When the northern poles are closest, a burst is launched northward, and vice versa for the southern poles. For every stellar period, two bursts are thus launched in sequence in opposite directions. 

We show a snapshot of our simulation in Figure \ref{fig:slice2d}, at $t = 3.42\,P_*$, when the bubble and trailing current sheet have formed and are heading northward. The main panel (b) shows a 3D volume rendering of the current density, with the two stars shown as light blue spheres with magnetic field lines traced from their surfaces. The bright, orange structure in the middle of the panel above the two stars is the trailing current sheet. The curved tube above it is the bubble. The right panel (c) shows 2D slices of the current density (top) and density (bottom) on a plane passing midway between the two stars, perpendicular to the current sheet. Field lines on this plane show the oppositely-oriented magnetic field on either side of the reconnecting sheet. The other sheet-like structures on either side of the stars are the equatorial current sheets of the individual stars. Panel (a) shows the evolution of the burst across five snapshots. An accompanying animation showcases the full dynamics.

Let us now go into further detail concerning the complex magnetic topology that gives rise to these periodic bursts. The following description will be for a burst headed northward; for a southward burst, the description is flipped.

At the moment when the northern poles are farthest apart, a significant fraction of polar field lines from the northern pole of one star does not open to infinity, but rather connects to the northern pole of the other star, forming `northern' loops with footpoints on either star. At the same time, since the southern poles are close together, a bundle of field lines from the southern pole of one star connects to the southern pole of the other star, forming a flux tube. As the southern poles draw farther apart as the stars rotate, this `southern' flux tube which sits between the two stars near the $z$ midplane is stretched between the stars until it becomes perpendicular to the overlying arcade of northern field lines. This configuration is strikingly similar to the standard picture of the conditions that give rise to CMEs on the Sun's surface \citep{cmes}. Analogously to CMEs, the flux tube then begins to rise upwards, pushing against the overlying northern field lines. This moment is shown in the left panel of Figure \ref{fig:fieldlines}, at $t=3.24\,P_*$. Similarly to Figure \ref{fig:slice2d}, we show a volume rendering of the current density. However, in Figure \ref{fig:fieldlines}, a different selection of field lines is plotted: the overlying northern arcade in blue, and the rising southern flux tube in green. We also show two different views with respect to the forming current sheet and bubble, one face-on (top row) and one side-on (bottom row). The right-hand column shows a slightly later time $t = 3.42\,P_*$ (the same time as in Figure \ref{fig:slice2d}), when the bubble, composed of green southern field lines, has risen, with the blue field lines reconnecting behind it. We include an animation of Figure \ref{fig:fieldlines} showing the full burst from formation to ejection.

In the canonical CME picture, reconnection of the overlying arcade of northern field lines behind the flux tube results in a bubble of magnetic flux entirely detached from the stars, as well as a `post-eruption arcade' connecting the two northern poles. This is the physical picture described in \cite{most2020apjl} for perfectly counter-aligned stars, for example. However, we find that the inclination of the stars complicates this picture. Indeed, in order to form a perfectly detached loop, an overlying field line must reconnect precisely with itself. When the stars are inclined, however, the magnetic topology is no longer as simple, and these overlying field lines can instead first reconnect with field lines within the flux tube itself, or with the closed zone of the other star. This process does not change the overall energetics of the system, since ultimately the dissipated energy in the sheet is set by the available free energy stored in the twist. There is, however, a topological difference: rather than detached magnetic loops circling around the ejected bubble, we observe helical ribbons winding around the flux tube. These helical ribbons are still connected to the poles of the stars as the bubble flies outwards, and they carry current. Some of these helical field lines can be seen looping around the flux tube in the top right-hand panel of Figure \ref{fig:fieldlines}, in green.

An alternative interpretation of these helical ribbons around the tube is the following. The footpoints on the stars to which the southern flux tube field lines and the northern overlying field lines are anchored do not all lie in the same plane, as they do when the stars are perfectly counter-aligned, thus breaking planar symmetry. Following the terminology of \cite{wright1989}, used in the context of solar physics, there is thus an initial `mutual helicity' between the flux tube and the overlying field lines. As reconnection proceeds, seeing as total helicity is approximately conserved, this `mutual helicity' is converted into the `self-helicity' of the resulting helical ribbons that enrobe the rising flux tube. By this logic, any CME-like reconnection process that does not have perfect planar symmetry (as is likely to be the case in reality) would result in these helical ribbons surrounding the escaping bubble. Therefore, we argue that the creation of these current-carrying helical ribbons is not unique to the stellar inclinations chosen in our particular simulation, but is rather a generic result for these anti-aligned binary systems, if the magnetic moments of the stars are not perfectly counter-aligned.

After the bubble and trailing current sheet depart, the whole process repeats on the southern side of the binary. We now turn to the magnetic dissipation and particle acceleration occurring in the trailing current sheet during each of these eruptions.

\subsection{Particle Acceleration} \label{subsec:particleaccel}

\begin{figure}[t] % or [!t] or [h] depending on where you want it
    \centering
    % 0.8 * page width in a two-column document
    \includegraphics[width=1\linewidth]{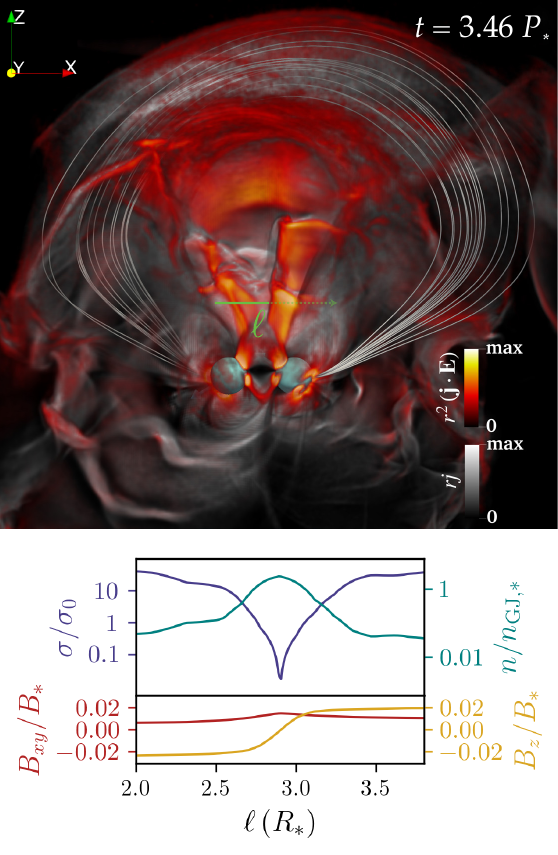}  
    \caption{\emph{Top}: Volume rendering of the current density $j$ (compensated by $r$) (grayscale), overlaid with the dissipation rate $\bm{j}\cdot\bm{E}$ (compensated by $r^2$) (red), with field lines in white showing the flux tube. \emph{Bottom}: 1D profile of $\sigma/\sigma_0$ (dark blue), $n/n_{\mathrm{GJ,*}}$ (teal), $B_{xy}/B_*$ (red), and $B_z/B_*$ (yellow) along a line $\ell$ piercing through the current sheet (see light green arrow in top panel). The reconnecting field $B_z$ begins to reverse when $\sigma \sim \sigma_0$, while the guide field $B_{xy}$ remains roughly constant.}
    \label{fig:diss}
\end{figure}

The trailing vertical current sheet is the dominant site of magnetic dissipation in the system. We show this in the top panel of Figure \ref{fig:diss}, which plots a volume rendering of the current density (in gray, to show the full structure of the magnetosphere) at $t = 3.46\,P_*$, overlaid with the dissipation rate $\bm{j} \cdot \bm{E}$ (red). There are traces of dissipation around the bubble, due to reconnection between the flux tube field lines (shown here in white) and the helical ribbons enwrapping them (see right column of Figure \ref{fig:fieldlines}), but the bulk of the dissipation occurs in the trailing vertical sheet. The dissipation in the sheet at this time is not completely uniform because, by this point in the sheet's evolution, the sheet is kinking in the $xy$ plane.

\begin{figure}[t] % or [!t] or [h] depending on where you want it
    \centering
    % 0.8 * page width in a two-column document
    \includegraphics[width=1\linewidth]{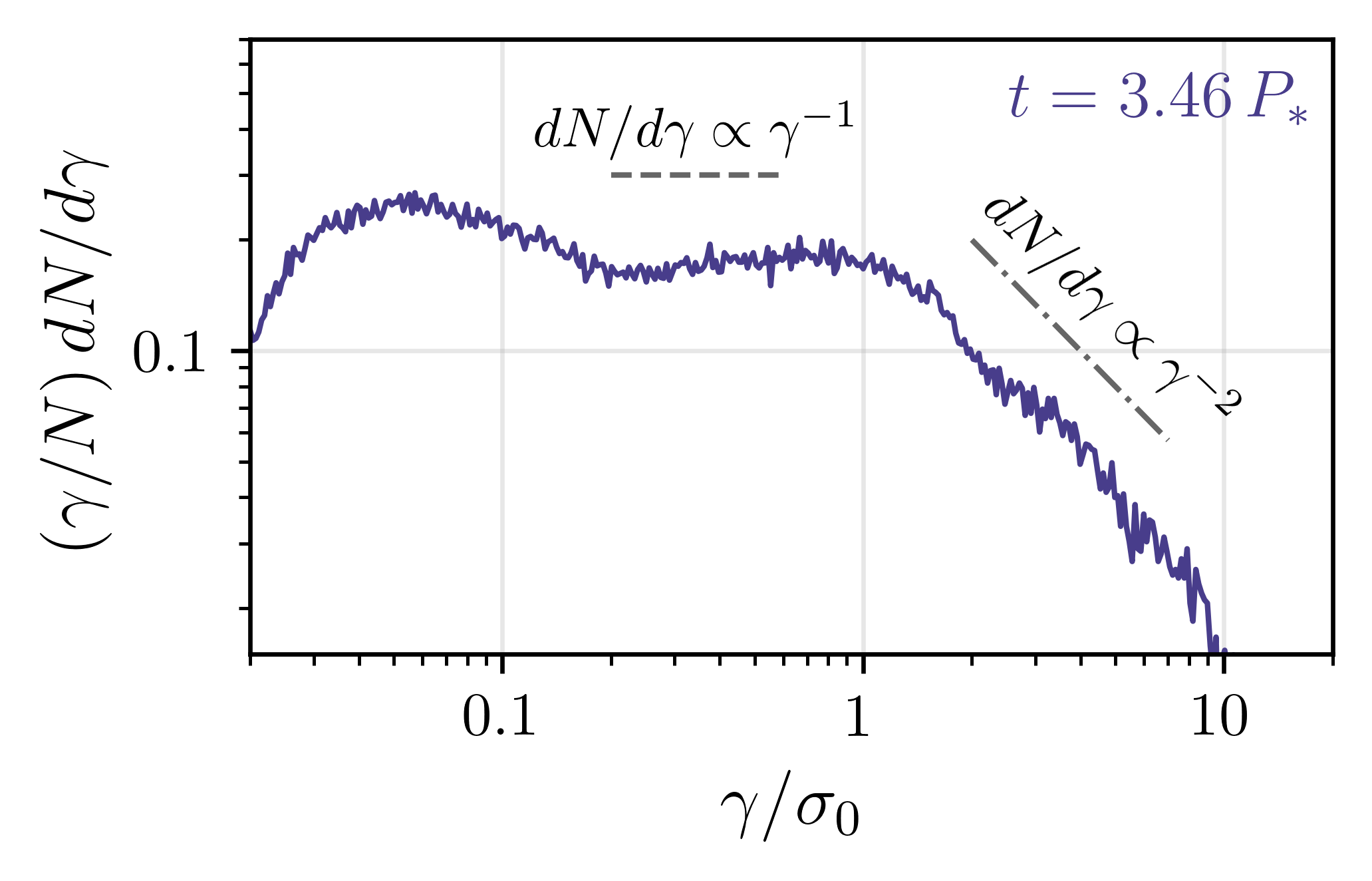}
    \caption{Particle spectrum of the trailing current sheet at $t = 3.46\,P_*$. Particle energies follow $dN/d\gamma\propto\gamma^{-1}$ before steepening at $\gamma\sim\sigma_0$ to $dN/d\gamma\propto \gamma^{-2}$, as expected from previous simulations of relativistic reconnection.}
    \label{fig:spectrum}
\end{figure}

A 1D profile of magnetization, density, $B_z$, and $B_{xy}$ along a line normal to the sheet is shown in the bottom panel of Figure \ref{fig:diss}. These components of $\bm{B}$ reveal the geometry of reconnection in the vertical sheet. The reversing component of $\bm{B}$ is $B_0 \equiv B_z$, while $B_g \equiv B_{xy}$ constitutes the non-reversing guide field. This guide field arises from the inclination of the stars, since reconnecting polar field lines are inclined with respect to the $z$ direction (i.e., the spin axis). We observe that $B_g \sim 0.3B_0$. We note that the value of the magnetization when $B_z$ starts to reverse is $\sigma\sim\sigma_0$, which in our simulation units is $\sigma_0\sim100$. The spectrum of particles contained within the current sheet at $t=3.46\,P_*$ is shown in Figure \ref{fig:spectrum}. The particles follow $dN/d\gamma\propto \gamma^{-1}$ before a break at $\gamma\sim\sigma_0$, at which point the power-law steepens to $dN/d\gamma\propto \gamma^{-2}$, consistent with previous kinetic studies of relativistic reconnection \citep[e.g.,][]{sironi2014}. We note, however, that in our global simulation, when observing the trajectories of particles with $\gamma \gtrsim 3\sigma$, they do not gain energy by traveling across the reconnection layer and probing the large fields in the upstream region, as has been shown in local 3D reconnection simulations \citep[see, e.g.,][]{zhang2021}. Rather, there are regions of the dynamically evolving sheet that have $\sigma\sim1-3\,\sigma_0$ and are thus able to accelerate particles beyond $\gamma\sim3\sigma_0$, up to $\gamma\sim10\sigma_0$.

We also calculate the bulk Lorentz factor of particles in the sheet. In the presence of a guide field, plasmoids are expected to reach a maximal bulk $\Gamma$ of
\begin{equation} \label{eqn:gammamax}
    \Gamma_\mathrm{max} \sim \sqrt{\frac{1+\sigma_g + \sigma_0}{1+\sigma_g}}
\end{equation}
where $\sigma_g$ is the magnetization parameter associated with the guide field instead of the reconnecting field \citep{liu2015}. As previously mentioned, in our simulation, $B_g\sim0.3B_0$, which means $\sigma_g\sim 0.09\sigma_0$, and so $\Gamma_\mathrm{max} \sim \sqrt{\frac{1.09\sigma_0}{0.09\sigma_0}} \sim 3.5$, for $\sigma_0 \gg 1$. Indeed, when calculating the bulk $\Gamma$ of our particles directly from the simulation, we obtain $\Gamma \gtrsim 2$, consistent with expectations. This estimate of the plasmoids' bulk Lorentz factor will become relevant for our prediction of coherent radio emission from plasmoid mergers in the trailing current sheet, detailed in subsection \ref{subsec:radio}.

Having established that the vertical current sheet efficiently dissipates magnetic energy and accelerates particles, we now turn to the potential for observing electromagnetic emission from this sheet.

\section{Observational implications}

\subsection{Energetics} \label{subsec:energetics}

In order to predict observable signatures, we must first estimate the energetics of the system, i.e., the outgoing Poynting flux carried by the bubble and the dissipated energy in the sheet. We integrate the radial Poynting flux, \mbox{$(4\pi/c)S_r = (\bm{E}\times\bm{B})_r$}, over a sphere of radius $\sim12.4\,R_*$ (approximately half the width of our simulation box), and the energy dissipation rate, $\bm{j}\cdot\bm{E}$, within that sphere, excluding regions close to the stars. Both the outgoing Poynting flux and the total dissipation are normalized to the spin-down luminosity of an individual star, $L_0 = (\mu^2 \Omega_*^4/c^3)(1+\sin^2\alpha)$, where $\mu = B_\mathrm{surf}R_*^3$ is the magnetic dipole moment, $B_\mathrm{surf}$ is the equatorial surface magnetic field strength, $R_*$ is the stellar radius, $\Omega_*$ is the stellar spin frequency, and $\alpha$ is the inclination angle \citep{spitkovsky2006}. We find that the outgoing Poynting flux peaks twice per stellar period, as expected, when the bubble and trailing current sheet pass through the spherical surface. The peak outgoing electromagnetic luminosity reaches  $\sim50\,L_0$, while the total dissipation in the current sheet peaks at $\sim10\,L_0$.

To translate these into physical units, we note that if $R_*\sim13\,\mathrm{km}$, then the orbital separation in our simulation is $a\sim45.5\,\mathrm{km}$, corresponding to an orbital period of $P = 2\pi\sqrt{a^3/(2GM_*)} \sim 3.2\,\mathrm{ms}$, where $M_* = 1.4\,M_\odot$ is the neutron star mass. For this orbital period, an inclination of $60^\circ$, and a typical surface magnetic field strength of $B_\mathrm{surf} = 10^{12}\,\mathrm{G}$, we obtain $L_0\sim5\times10^{42}\,\mathrm{erg/s}$. The peak outgoing electromagnetic luminosity is then $\sim 3\times10^{44}\,\mathrm{erg/s}$, and the peak dissipated luminosity is $\sim5\times10^{43}\,\mathrm{erg/s}$. These values are roughly consistent with results from force-free simulations of a similar configuration to ours by \cite{most2020apjl, most2022mnras}, in which orbital motion rather than stellar spin provides the twist.

Now, let us briefly describe the physics behind the outgoing Poynting flux carried by the bubble, which will allow us to see how the outgoing electromagnetic luminosity scales with orbital separation. Because the stars are intermittently magnetically tethered to each other, twist can accumulate between them, allowing $B_\phi$ to grow to $\sim B_\mathrm{pol}$ at the midpoint $a/2$. In a standard isolated pulsar, $B_\phi\sim B_\mathrm{pol}$ is only reached at $\sim R_\mathrm{LC}$ \citep[for a relevant discussion of compressed magnetospheres, see][]{zhong2024}. Following the standard spin-down derivation for an aligned rotator \citep[see, e.g.,][]{philippov2022review}, and evaluating the integrated Poynting flux not over a full sphere at $a/2$, but over the solid angle of the bubble, which is roughly $\sim 1/8 (4\pi(a/2)^2)$, we find
\begin{equation}
    L = \frac{\mu^2\Omega}{8(a/2)^3}.
\end{equation}

Here, the twist can be provided either by the individual spins of the stars $\Omega_*$ or $\Omega_\mathrm{orb}$. If $\Omega = \Omega_\mathrm{orb}$, the enhancement is then $L/L_0 \sim 1/8\,(R_\mathrm{LC}/(a/2))^3\sim1/8\,(c/v_\phi)^3$, where $v_\phi \sim \Omega_\mathrm{orb}a/2$ is the corotation velocity at the midpoint. At a separation of $a\sim 45.5\,\mathrm{km}$, $v_\phi\sim0.15c$, yielding $L/L_0\sim37$, consistent with our simulation result of $L/L_0\sim50$.

In terms of scalings, for non-spinning stars \mbox{$\Omega_\mathrm{orb} \propto a^{-3/2}$}, and so $L\propto a^{-9/2}$. If the twist is instead provided by the stellar spin, then $\Omega$ is independent of separation, and $L\propto a^{-3}$. We assume that the dissipated luminosity follows these same scalings.\footnote{If the rotation is sufficiently slow, then it is possible that the lifetime of the trailing current sheet is determined not by the timescale of the twist, as is assumed above, but rather the reconnection timescale, changing the scaling of the dissipated luminosity in the sheet to $L\propto a^{-4}$ (see \cite{most2020apjl}). For the orbital separation shown in our simulation, the reconnection timescale is $t_{\mathrm{rec}}\sim2a/v_\mathrm{rec}\sim3\,\mathrm{ms}$, where $v_\mathrm{rec}\sim0.1c$ is the reconnection speed. This reconnection timescale is thus comparable with the orbital period at this separation ($P\sim3.2\,\mathrm{ms}$). For the purposes of the observational implications presented in this Letter, however, this difference in scaling has no significant impact.} In the estimates that follow, we also assume a transition between the two regimes when $\Omega_\mathrm{orb}\sim\Omega_*$, which, for a typical pulsar spin period of $0.1\,\mathrm{s}$, corresponds to a separation of $a\sim450\,\mathrm{km}$. In other words, we adopt the steeper scaling of $L\propto a^{-9/2}$ for separations $a\lesssim450\,\mathrm{km}$, and the shallower scaling of $L\propto a^{-3}$ for separations $a\gtrsim450\,\mathrm{km}$. We emphasize that the $L\propto a^{-9/2}$ drop-off for small separations may be too steep, if the stellar spins are able to provide twist as well.

\begin{figure*}[t] % or [!t] or [h] depending on where you want it
    \centering
    % 0.8 * page width in a two-column document
    \includegraphics[width=0.8\paperwidth]{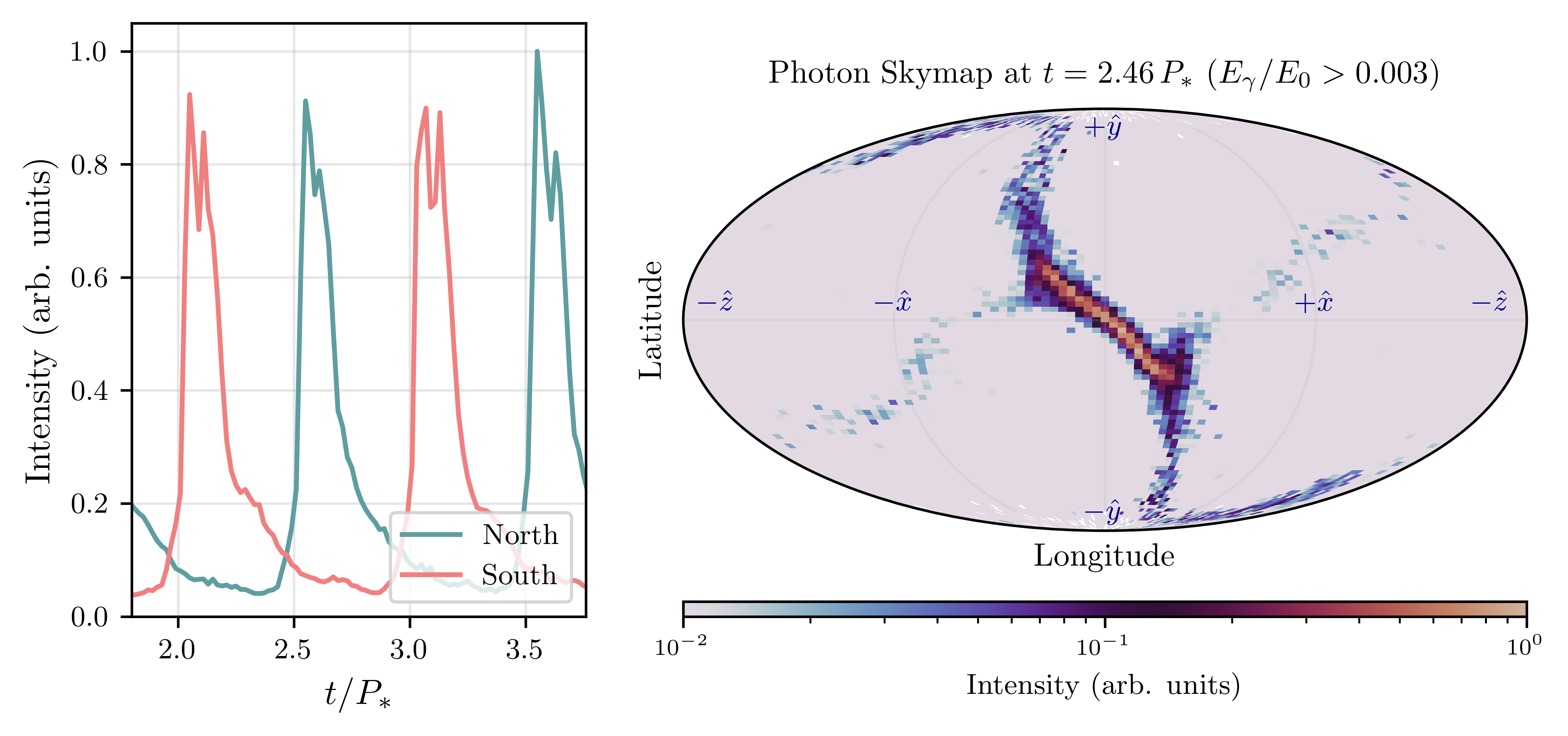}
    \caption{\emph{Left}: Synchrotron lightcurves over two stellar periods for a northward observer (blue) and a southward observer (red). Bursts occur twice per period, with duration $\Delta t\sim1/6 \,P_*.$ \emph{Right}: Intensity skymap of high-energy synchrotron photons at $t = 2.46\,P_*$ on a spherical surface of radius $R\sim12.4R_*$, shown in the Mollweide projection, where the center is the $+z$ axis. Photon energies are normalized to $E_0\propto\epsilon_\mathrm{sync}(\sigma/\gamma_\mathrm{rad})$. The radiation reflects the geometry of the trailing current sheet which is flying out in the $+z$ direction, concentrated into a solid angle $\Delta\Omega\sim0.29\,\mathrm{sr}$.}
    \label{fig:skymap}
\end{figure*}

\subsection{High-energy signal} \label{subsec:gammaray}

    Since the magnetic field in the flare during the final orbits before merger is likely to be very high, we expect the reconnection in the vertical sheet to proceed in the strong cooling regime, and can thus reasonably assume that the dissipated energy is immediately converted into radiation \citep{beloborodov2021}. However, the large magnetic compactness of the flare of order $l_B \sim 10^5 -10^6$ \citep{beloborodov2021, metzger2016} at these small orbital separations means that high-energy photons pair produce in the upstream of the reconnection layer, creating an optically thick coat of pair plasma that prevents nonthermal high-energy radiation from escaping. \cite{beloborodov2021} showed that the resulting signal in this case would be an X-ray burst with a thermal spectrum extending to $E_c \sim 0.1\,m_e c^2$ with $L\gtrsim10^{42}\,\mathrm{erg/s}$ for $B_\mathrm{surf} \gtrsim 10^{13}\,\mathrm{G}$, which is  detectable only for nearby mergers with a sensitive X-ray telescope. Our global simulation does not include the radiative physics needed to self-consistently compute the radiation spectrum in this ultra-high-compactness regime, which has not yet been explored even in local kinetic reconnection simulations. However, the optical depth to pair production, $\tau_{\gamma\gamma}$, should drop at larger orbital separations $a$. This is because the distance from the binary at which the flare occurs is $\propto a$, and so \mbox{$B_\mathrm{flare} \propto a^{-3}$}. Thus, there is a time before merger at which $\tau_{\gamma\gamma}$ is sufficiently low that the reconnecting sheet would be optically thin, allowing nonthermal gamma-ray photons produced in the sheet to escape. 

    We now characterize the optically thin radiation produced by the current sheet, before estimating the orbital separations at which it could escape. We post-process our particle data to produce synchrotron radiation skymaps and lightcurves, following the method of \cite{cerutti2012, cerutti2016}. Namely, we assume that particles emit \mbox{(macro-)photons} with critical frequency ${\hbar\omega_c = \epsilon_\mathrm{syn} (\gamma/\gamma_\mathrm{rad})^2 \, \chi_R}$, where $\gamma$ is the Lorentz factor of the particle and $\chi_R$ is the normalized effective magnetic field perpendicular to the particle's motion in its rest frame. Here, $\gamma_\mathrm{rad} = \sqrt{3e\beta_\mathrm{rec}B/(2r_e^2 B^2)}$ is the Lorentz factor for which the energy gain from reconnection balances synchrotron losses \citep{mahlmann2022, hakobyan2023pulsar}, where $r_e$ is the classical electron radius and ${\beta_\mathrm{rec} \sim 0.1}$ is the reconnection rate, and ${\epsilon_\mathrm{syn} \sim \beta_\mathrm{rec}(9/4)m_ec^2\alpha^{-1} \sim 16\,\mathrm{MeV}}$ is the synchrotron burn-off energy, where $\alpha$ is the fine-structure constant \citep{uzdensky2011}. The full synchrotron spectrum is then determined by the classical synchrotron kernel at this critical frequency. In the environment of the flare, $\gamma_\mathrm{rad} \lesssim \sigma$, placing us in the strong cooling regime where particles radiate immediately after being accelerated. Because our simulation does not constrain the true pair multiplicity (which may be higher than in our simulation), however, we do not have sufficient constraints on $\sigma$ to map our simulation energies to physical units. This being said, our results for the lightcurve morphology and angular distribution of the radiation are independent of $\sigma$, as long as $\sigma \gg \gamma_\mathrm{rad}$.

    In Figure \ref{fig:skymap}, we show the synchrotron lightcurve on the left, and the angular distribution of the radiation at $t = 2.46\,P_*$ on the right. To compute the lightcurve shown in blue, we place a detection plane at the northern boundary of our simulation domain, infinite in $x$ and $y$; for the lightcurve shown in red, the detection plane is instead placed at the southern boundary. We also only keep photons that are within a cone of half-angle $60^\circ$ from the $z$ axis. At each observer time, we sum the energy flux from all photons crossing the plane, accounting for the light travel time from each photon's emission point to the detector. The lightcurve shows periodic bursts of radiation, consistent with the periodic eruptions described in Section \ref{sec:simresults}. Each burst has a duration of $\Delta t\sim1/6\,P_*$.

    Importantly, the radiation intensity of high-energy photons has an anisotropic distribution across the sky, as seen in the right-hand panel. We plot our skymap such that the $+z$ direction is in the center of the projection. Looking at the streak of orange-yellow, we observe that the radiation is not isotropic, but rather geometrically beamed along the trailing current sheet. This is because the accelerated particles themselves are not isotropic, but rather are preferentially accelerated up the sheet, in the direction of the upstream magnetic field. The blue extensions to the main streak are photons from the equatorial current sheets of the stars.

    The effect of this anisotropic emission is twofold. First, an observer would infer an isotropic-equivalent luminosity that is larger than our estimate of the dissipated luminosity, i.e., $L_\mathrm{obs}\sim  (4\pi/\Delta\Omega)\,L_\mathrm{diss}$, where $\Delta\Omega$ is the solid angle in which the radiation is being concentrated. We can directly estimate the solid angle within which half of the intensity is contained to be $\Delta\Omega_\mathrm{50\%}\sim0.29 \,\mathrm{sr}$, yielding a luminosity boost of $\sim40$. We note that this solid angle should depend on the opening angle of the sheet, which is determined by the reconnection rate and not on the scale separation or resolution used in our simulation.
    
    Second, this anisotropy affects our estimate of the optical depth to pair production, $\tau_{\gamma\gamma}$, since collisions between photons are now less likely due to their preferential beaming. The interaction rate between photons, and thus $\tau_{\gamma\gamma}$, now must include a factor of $(1-\cos(\theta))$ where $\theta$ is the average angle between photons. Conservatively, we estimate this angle to be $\sim 60^\circ$, i.e., the angular extent of the orange-yellow streak in Figure \ref{fig:skymap}. We note that lower energy photons are more isotropic than the high-energy photons shown in Figure \ref{fig:skymap}, which, if they were to pair produce with high-energy photons, would reduce the significance of this effect.
    
    We now estimate $\tau_{\gamma\gamma}$ following the methodology of \cite{hakobyan2023}. The characteristic optical depth for photon-photon interactions of average energy $\epsilon_\gamma$ is $\tau_{\gamma\gamma} \sim \sigma_{\gamma\gamma}n_\gamma S\,(1-\cos \theta)$, where $n_\gamma$ is the number of photons of energy $\epsilon_\gamma$, $S$ is the size of the interaction region, and $\sigma_{\gamma\gamma}$ is the Breit-Wheeler cross-section, which at its maximum is $\sim 0.25\,\sigma_T$ \citep[see, e.g.,][]{gould1967}. 

    The photon number density can be estimated as \mbox{$n_\gamma \sim U_\mathrm{rad}^\mathrm{sync}/\epsilon_\gamma \sim B^2\beta_\mathrm{rec}/(4\pi\epsilon_\gamma)$}, where $U_\mathrm{rad}^\mathrm{sync}$ is the radiation energy density produced by dissipation of the magnetic field. Since the rate of pair production is controlled by the highest-energy photons, we take $\epsilon_\gamma \sim \epsilon_\mathrm{sync} \sim\,16\,\mathrm{MeV}$.
    Taking the size of the reconnecting sheet to be $S\sim 2a$ and the flare field strength to thus be $B_\mathrm{flare} \sim B_\mathrm{surf}\left(2a/R_*\right)^{-3}$ (assuming an unperturbed dipolar field scaling), the optical depth becomes
\begin{equation}
    \begin{split}
        \tau_{\gamma\gamma}
        &\approx 2000 \left(\frac{a}{45.5\,\mathrm{km}}\right)^{-5}\left(\frac{B_\mathrm{surf}}{10^{12}\,\mathrm{G}}\right)^2
    \end{split}
\end{equation}

\noindent    for $R_* = 13\,\mathrm{km}$. Setting $\tau_{\gamma\gamma}\sim1$ gives the orbital separation, and equivalently the time before merger, at which nonthermal gamma-ray photons can escape.

\begin{figure}[t] % or [!t] or [h] depending on where you want it
    \centering
    % 0.8 * page width in a two-column document
    \includegraphics[width=1\linewidth]{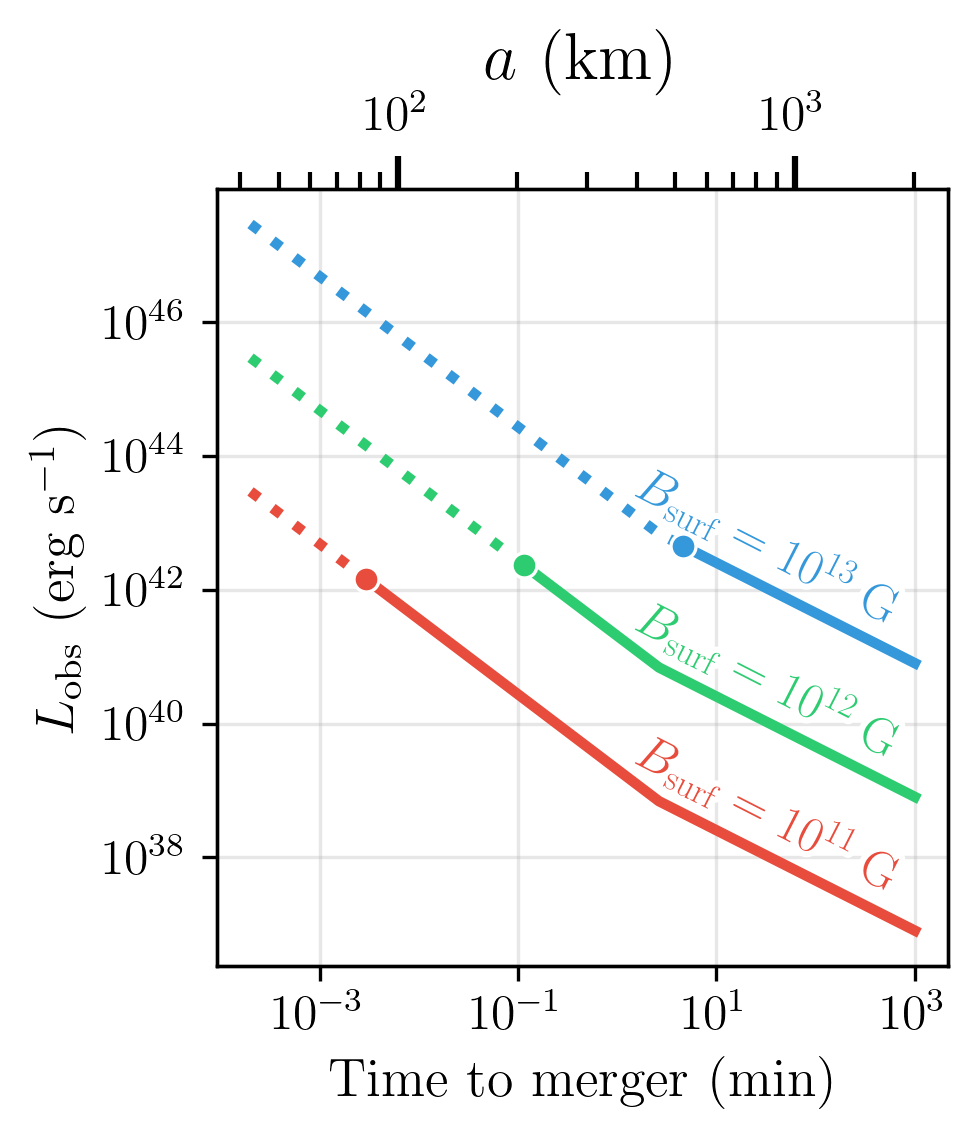}
    \caption{Observed gamma-ray luminosity for a head-on observer as a function of time to merger (bottom) or orbital separation (top), for three different surface magnetic field strengths ($B_\mathrm{surf} = 10^{11}\,\mathrm{G},\, 10^{12}\,\mathrm{G},\, 10^{13}\,\mathrm{G}$ in red, green, and blue, respectively). The dots mark the times at which the sheet ceases to be optically thin to pair production ($\tau_{\gamma\gamma}\sim1$). To the right of the dots (solid curves), the nonthermal gamma-ray burst, peaking at $\sim16\,\mathrm{MeV}$, would be able to escape. To the left of the dots (dotted curves), the emission becomes optically thick and is Comptonized into a thermal X-ray signal of lower luminosity than $L_\mathrm{obs}$, since the emission would no longer be geometrically beamed. The break at $a\sim450\,\mathrm{km}$ marks the transition from twist provided by stellar spin ($L\propto a^{-3}$) to twist provided by the orbital motion ($L\propto a^{-9/2}$) for a typical stellar period of $P\sim0.1\,\mathrm{s}$ (see subsection \ref{subsec:energetics}).}
    \label{fig:gammaemission}
\end{figure}

Combining this optical depth estimate with the luminosity scalings from subsection \ref{subsec:energetics}, Figure \ref{fig:gammaemission} shows the observed gamma-ray luminosity as a function of time to merger\footnote{Note that we calculate the time to merger as \mbox{$t = 5c^5a^4/(256G^3m_1m_2(m_1+m_2))$}, where \mbox{$m_1 = m_2 = 1.4\,M_\mathrm{sun}$} are the neutron star masses.} (bottom axis) or orbital separation (top axis) for three different surface magnetic field strengths. For each of the surface magnetic field curves, the dots correspond to the times at which $B_\mathrm{flare}$ is sufficiently low that $\tau_{\gamma\gamma} \sim 1$, and thus gamma-ray photons can freely escape from the reconnecting sheet. For a surface field of $B_\mathrm{surf} \sim 10^{12}\,\mathrm{G}$, for example, we expect the sheet to remain optically thin until a few seconds before merger, with a peak observed luminosity of $\gtrsim10^{42}\,\mathrm{erg/s}$.
After this point, the nonthermal gamma-ray photons efficiently pair produce, creating an optically thick layer of pair plasma around the sheet that Comptonizes the emission into a thermal X-ray signal with observed luminosities lower than the nonthermal gamma-ray signal even at smaller orbital separations \citep{beloborodov2021}.

As has been shown in previous studies of radiative reconnection, the peak energy of this escaping nonthermal synchrotron emission is simply set by the synchrotron burn-off limit, $E_\mathrm{peak} \sim 16\,\mathrm{MeV}$ for $\beta_\mathrm{rec}\sim 0.1$. In the strong cooling regime, the spectral cutoff is then determined by $\sigma$, i.e., $E_\mathrm{cut} \sim E_\mathrm{peak}(\sigma/\gamma_\mathrm{rad})$ \citep[see, e.g.,][]{chernoglazov2023, hakobyan2023, hakobyan2023pulsar}. Since, as previously mentioned, we do not have constraints on $\sigma$ in these premerger binary systems, it is challenging to predict the spectral cutoff of our resulting signal. It could, for example, plausibly extend into the $\mathrm{GeV}$ range. We note that our assumption of strong cooling ($\gamma_\mathrm{rad}\ll\sigma$) breaks down at sufficiently large orbital separations, as $B_\mathrm{flare}$ drops and the radiation-reaction-limited Lorentz factor approaches the magnetization ($\gamma_\mathrm{rad}\sim\sigma)$. However, even for relatively high multiplicities of $\kappa \sim 10^{4-5}$, $\gamma_\mathrm{rad} \lesssim \sigma$ up to orbital separations of several thousand $\mathrm{km}$, thus validating our assumption of strong cooling.

For sub-magnetar surface fields, detecting such a signal would require a nearby merger. A burst with luminosity $\sim10^{42}\,\mathrm{erg/s}$ peaking at $\sim16\,\mathrm{MeV}$ would be observable by Fermi-GBM \citep{fermi2009} only at close distances, and the relatively small solid angle of emission further reduces the detection probability. One of the stars having a magnetar-strength field or a millisecond spin period, or the presence of significant relativistic beaming effects, would improve this prospect.

\begin{figure}[t] % or [!t] or [h] depending on where you want it
    \centering
    % 0.8 * page width in a two-column document
    \includegraphics[width=\columnwidth]{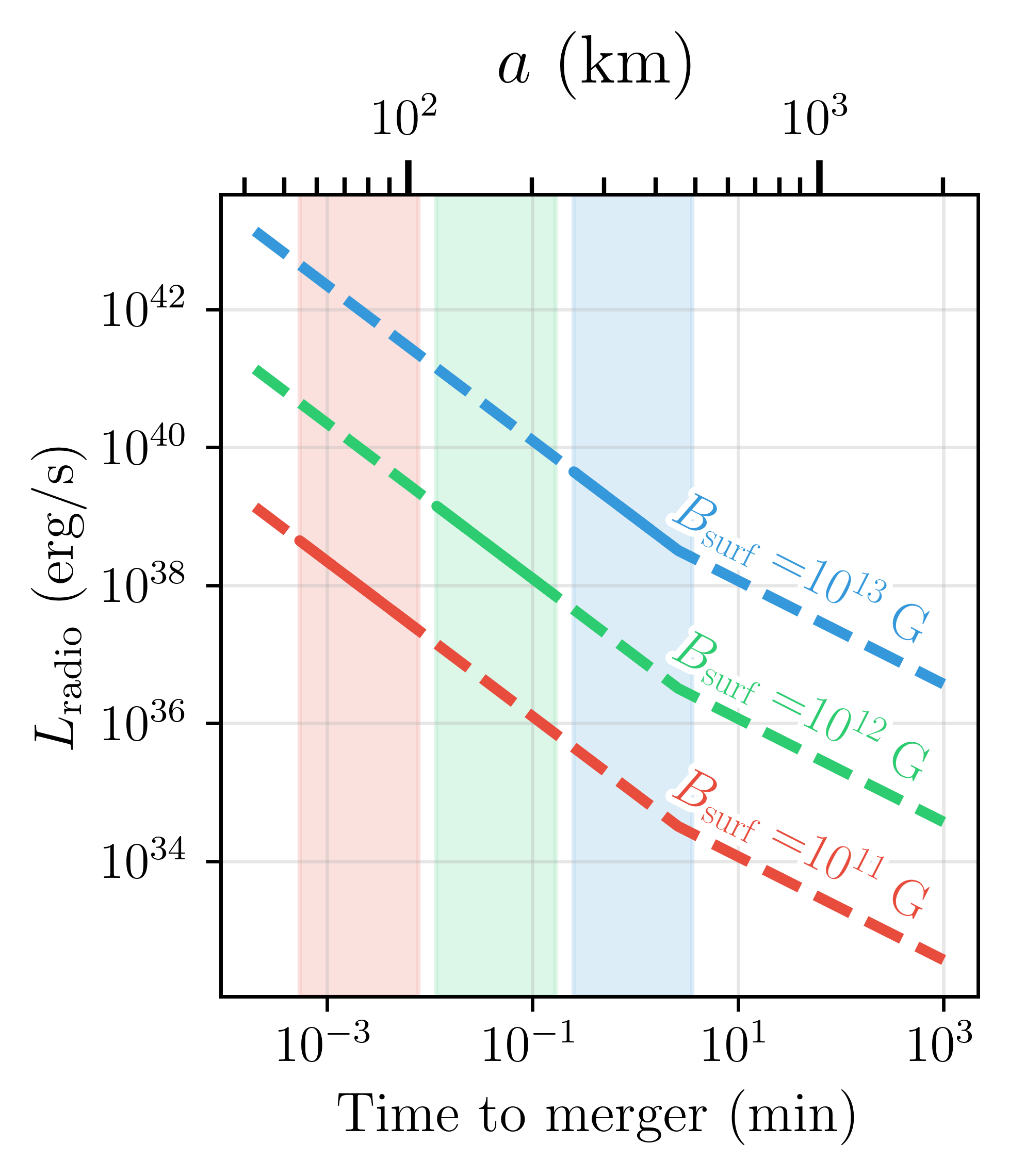}
    \caption{Radio luminosity as a function of time to merger (bottom) or orbital separation (top), for three different surface magnetic field strengths ($B_\mathrm{surf} = 10^{11}\,\mathrm{G},\, 10^{12}\,\mathrm{G},\, 10^{13}\,\mathrm{G}$ in red, green, and blue, respectively). The shaded regions correspond to times in which plasmoid mergers in the trailing current sheet would emit coherent radio waves in the frequency range $400\,\mathrm{MHz}-8\,\mathrm{GHz}$, for each surface magnetic field strength. The luminosity curves are dashed when the emitted frequency is outside of that frequency range; to the left of the shaded regions, the frequency is higher than $8\,\mathrm{GHz}$, and to the right, lower than $400\,\mathrm{MHz}$. The break at $a\sim450\,\mathrm{km}$ marks the transition from twist provided by stellar spin ($L\propto a^{-3}$) to twist provided by the orbital motion ($L\propto a^{-9/2}$) for a typical stellar period of $P\sim0.1\,\mathrm{s}$ (see subsection \ref{subsec:energetics}).}
    \label{fig:radioemission}
\end{figure}

\subsection{Coherent radio signal} \label{subsec:radio}

Beyond the high-energy emission discussed above, the trailing current sheet could also produce coherent radio emission, which as we will see in this subsection is a more promising observational prospect. Indeed, \cite{philippov2019nanoshot} and \cite{Lyubarsky2019} showed that merging plasmoids in a reconnecting current sheet can source coherent electromagnetic waves, offering an explanation for coherent radio emission from beyond the light cylinder observed in some pulsars. This mechanism has recently been extended to magnetars, in which a flare from the inner magnetosphere flies out and compresses the current sheet beyond the light cylinder, potentially producing fast radio bursts \citep{lyubarsky2020, mahlmann2022}. Building on this picture, \cite{mostPRL} also proposed that a flare from a pre-merger binary neutron star system, similar to the one discussed in this Letter, could compress the orbital current sheet of the binary, producing a high-frequency fast radio transient.

Here, however, we focus on radio emission from the trailing current sheet itself, most analogous to work on isolated pulsar current sheets. Following \cite{philippov2019nanoshot, Lyubarsky2019, lyubarsky2020, mahlmann2022}, the characteristic frequency of coherent radiation from plasmoid mergers in the lab frame is $\nu = \Gamma \nu'$, where $\nu'$ is the characteristic frequency in the frame of the sheet. The frequency $\nu'$ arising from plasmoid mergers is $\nu' = c/(2\pi\xi H')$, where $\xi \sim 10-100$ is the ratio of the plasmoid size to the current sheet thickness $H'$. The sheet thickness scales as $H' \sim \zeta r'_L$, where $r_L' = m_e c^2 \langle\gamma'\rangle/eB_\mathrm{flare}$ is the typical Larmor radius of particles in the sheet, and so $\zeta\sim1-10$ is the number of Larmor radii that fit into the sheet. In our pre-merger magnetosphere, synchrotron cooling is strong, since $\gamma_\mathrm{rad} \ll \sigma$. We can therefore estimate that $\langle\gamma'\rangle \sim \gamma_\mathrm{rad}$, yielding

\begin{equation} \label{eqn:radio}
    \nu = \frac{\Gamma}{2\pi\xi\zeta}\left(\frac{2r_e}{3\beta_\mathrm{rec}c}\right)^{1/2} \omega_B^{3/2},
\end{equation}
where $\omega_B = eB_\mathrm{flare}/(m_ec)$ is the gyrofrequency.\footnote{We note that Equation \ref{eqn:radio} is identical to the expression derived in \cite{mahlmann2022}, except that here $\nu \propto \Gamma$ rather than $\nu \propto \Gamma^{-1/2}$, since our $B_\mathrm{flare}$ is already in the lab frame and does not need to be transformed into the frame of the compressed sheet.}

Using results from kinetic reconnection simulations in \cite{mahlmann2022}, $\xi\zeta\approx 90$, and the efficiency of converting dissipated luminosity into radio is \mbox{$f \sim 2\times10^{-3}$}. We furthermore use $\Gamma = 3.5$ (see subsection \ref{subsec:particleaccel} for a discussion of $\Gamma_\mathrm{max}$ of plasmoids in the sheet). Since the emitted frequency now depends only on $B_\mathrm{flare}$, which scales as $a^{-3}$, we can determine at which orbital separations and surface magnetic fields a coherent radio signal would fall within observable radio bands.

We present these estimates in Figure \ref{fig:radioemission}. Similarly to Figure \ref{fig:gammaemission}, the three colored lines represent the expected luminosity (here, in radio) for three different surface magnetic field strengths as a function of time to merger (bottom axis) or separation (top axis). The shaded regions then correspond to the separations that would yield a sufficiently low $B_\mathrm{flare}$ in order to produce radio emission in the range $0.4-8\,\mathrm{GHz}$, for each $B_\mathrm{surf}$. In other words, to the left of the shaded regions, since the separation between the stars is smaller, $B_\mathrm{flare}$ (and thus $L_\mathrm{radio}$) is higher, but the frequency of the emitted waves now exceeds $8\,\mathrm{GHz}$. We see that, depending on the surface magnetic field, FRB-like transients with $L\sim 10^{38-40}\,\mathrm{erg/s}$ could be produced minutes to seconds before merger. These bursts could show sub-ms variability. Successive bursts are expected to rise in frequency, reflective of the shrinking orbital separation, and also potentially shorten in duration if the twist is provided by the orbital motion.

To estimate a typical spectral flux density, consider a burst with radio luminosity $10^{39}\,\mathrm{erg/s}$ at $\sim6\,\mathrm{s}$ before merger. Assuming an emission bandwidth of $\sim300\,\mathrm{MHz}$ and a distance to the source of $\sim300\,\mathrm{Mpc}$, this corresponds to a spectral flux density of $\sim31\,\mathrm{mJy}$. For a $10\,\mathrm{ms}$ integration time (the orbital period of the binary being $\sim30\,\mathrm{ms}$ at this time), this is well above the noise floor for pointed instruments such as FAST \citep{fast2011} and the upcoming SKA-Mid \citep{ska2019}, and near the detection threshold for CHORD \citep{chord2019}, an upcoming wide-field survey instrument. The primary challenge therefore is not sensitivity but rather sky coverage, given the expected low rate of binary neutron star mergers of $7.6-250\,\mathrm{Gpc}^{-3}\,\mathrm{yr}^{-1}$\citep{ligo2025}. For untargeted detection, CHORD is thus best-suited to detect and localize these precursor bursts, combining wide instantaneous sky coverage with broad bandwidth. Even so, years of continuous monitoring may be required before catching a precursor burst. Since our model predicts multiple bursts per orbit, combining the signal from successive bursts could improve the overall signal-to-noise ratio, extending the effective detection horizon and thus the accessible volume.

A more promising strategy may be a multi-messenger approach. The O5 run of LIGO is expected to detect BNS mergers up to $\sim330\,\mathrm{Mpc}$ \citep{ligo2020}, and efforts are underway to improve sky localization in the final minute before merger \citep{magee2022,sachdev2020}. If sufficiently precise, say $\sim100\,\mathrm{deg}^2$, such localization could enable a wide-field radio instrument like ASKAP in the southern hemisphere \citep{askap2021}, with a $\sim30\,\mathrm{deg}^2$ field of view, or the upcoming DSA in the northern hemisphere \citep{hallinan2019, morsony2024}, with a $\sim10\,\mathrm{deg}^2$ field of view, to tile the localized region to search for a precursor burst in the final minute before merger. Our model is particularly well-suited to this approach, as it predicts multiple observable bursts in the final seconds before merger rather than a single event. A detection of one such precursor burst would sharply constrain the source position, aiding targeted follow-up by instruments with narrower fields of view, both in the radio and other electromagnetic bands. Alternatively, wide-field instruments such as CHORD or CHIME that continuously buffer raw data could be searched retrospectively at a known sky position and time once a merger is localized by LIGO post-coalescence, if the source was in the CHIME/CHORD field of view at the time of merger. Furthermore, gravitational wave detectors are most sensitive to face-on BNS systems \citep{chenligo2019}. This is advantageous for our model, as the coherent radio emission from the trailing current sheet is beamed roughly perpendicular to the orbital plane, if the stars' magnetizations are similar \citep[see][for a parameter survey of different stellar inclinations and magnetizations]{most2022mnras}.

These estimates come with several caveats. The guide field strength $B_g\sim0.3B_0$ found in our simulation depends on the stellar inclination and relative magnetization of the stars, and may also decrease with distance from the binary. A higher $\Gamma$ boosts the emitted waves to higher frequencies, thereby shifting the shaded regions in Figure \ref{fig:radioemission} to the right (i.e., to earlier times before merger). This would not necessarily reduce the observed luminosity, however, since the emission would be more tightly beamed, but it would change the time window during which the signal falls in the observable radio band. In addition, since the presence of a guide field likely affects the dynamics of plasmoid merging and thus the efficiency of converting dissipated luminosity into radio, our estimates of $L_\mathrm{radio}$, which rely on 2D simulations without guide field, may represent an upper bound if conditions for plasmoid merging are less favorable in 3D guide-field reconnection. Future studies of plasmoid merging in 3D, both with and without guide field, are required to further investigate the efficiency of coherent radio emission.

Importantly, the propagation of coherent radio emission through magnetospheric pair plasma is also the subject of active debate \citep[see, e.g.,][]{beloborodov2021frb, beloborodov2022frb, qu2022frb, golbraikh2023frb, mostPRL, lyutikov2024frb, sobacchi2024}. We note that unlike in a reconnection event happening deep within the closed zone of an isolated magnetar, the fast magnetosonic waves produced by plasmoid mergers in the trailing current sheet propagate on open field lines. This would prevent the nonlinear growth of the wave amplitude relative to the background field, facilitating the escape of the waves and conversion to electromagnetic radiation without steepening into monster shocks \citep{beloborodov2023, bernardi2025}.  In addition, the outgoing magnetic bubble preceding the sheet may modify the plasma environment along the escape path. A more detailed treatment of radio wave propagation in this pre-merger binary context, including for example potential nonlinear decay into Alfvén waves \citep{golbraikh2023frb}, is left to future work.

We note that other coherent radio mechanisms may also operate in this pre-merger environment, including a higher-frequency transient from the compression of the orbital current sheet by the outgoing bubble \citep{mostPRL}, coherent radio emission due to electromagnetic draping \citep{lyutikov2023, sharma2026}, and synchrotron maser emission from a bubble-driven shock \citep{metzger2019, sridhar2021}. In the latter case, the presence of current-carrying helical field lines enwrapping the bubble, as discussed in subsection \ref{subsec:global}, may be relevant for future modeling.

\section{Conclusions} \label{sec:conclusions}

We have presented the first global 3D kinetic simulations of a pre-merger neutron star binary. We find that, for anti-aligned dipoles, solar flare-like outbursts can occur multiple times per orbit, efficiently dissipating energy and accelerating particles in the reconnecting current sheet trailing the outgoing bubble. These accelerated particles immediately radiate away their energy, producing a nonthermal gamma-ray signal when the sheet is still optically thin to pair production, i.e., several minutes to seconds before merger. With modest luminosities of $L_\mathrm{obs}\gtrsim10^{42}\,\mathrm{erg/s}$, however, these gamma-ray precursors would only be detectable by Fermi-GBM for nearby mergers. More promisingly, plasmoid mergers in the trailing sheet could generate FRB-like transients in the $0.4-8\,\mathrm{GHz}$ range minutes to seconds before merger, with typical luminosity $L_\mathrm{radio}\sim10^{38-40}\,\mathrm{erg/s}$. Future work on the escape of these coherent radio waves from the pre-merger environment is required in order to further clarify the characteristics of these bursts.

Among proposed electromagnetic precursors to BNS mergers, the repeating coherent radio signal predicted here is uniquely suited to real-time follow-up of gravitational-wave early-warning alerts, as it occurs early enough and frequently enough to be searched for before coalescence. This could enable early electromagnetic localization of the source by instruments such as ASKAP, DSA, or SKA-mid, facilitating prompt multi-wavelength follow-up of the merger and its immediate aftermath.

\begin{acknowledgments}
The authors would like to thank Alexander Chernoglazov, Elias Most, Matthew Goodbred, and Robert Ewart for insightful discussions. This research was supported by NSF through Multimessenger Plasma Physics Center (MPPC, NSF grant PHY-2206607) and by Simons Foundation (grant MP-SCMPS-0000147). In addition, this research was supported in part by NSF grant PHY-2309135 to the Kavli Institute for Theoretical Physics (KITP). We  acknowledge the use of Princeton University's Research Computing resources for simulations in this paper.
\end{acknowledgments}

\bibliography{ns_precursor}{}
\bibliographystyle{aasjournalv7}

\end{document}